\documentclass{article}
\usepackage{arxiv}
\usepackage{hyperref}
\usepackage{latexsym}
\usepackage{url}
\usepackage[utf8]{inputenc}
\usepackage{booktabs}
\usepackage{amsmath}
\usepackage{amsfonts}
\usepackage{amssymb}
\usepackage[inline]{enumitem}
\usepackage{enumitem}
\usepackage{mathbbol}
\usepackage[T1]{fontenc}
\usepackage{mathtools}
\usepackage{multirow}
\usepackage{natbib}

\newcommand*\samethanks[1][\value{footnote}]{\footnotemark[#1]}
\newcommand{\ie}{\emph{i.e.}}
\newcommand{\eg}{\emph{e.g.}}
\newcommand{\wrt}{\emph{w.r.t. }}

\title{Incorporating Query Term Independence Assumption for Efficient Retrieval and Ranking using Deep Neural Networks}

\author{
  Bhaskar Mitra\thanks{Both authors contributed equally to this research.} \\
  Microsoft AI \& Research \\
  \texttt{bmitra@microsoft.com} \\
   \And
  Corby Rosset\samethanks \\
  Microsoft AI \& Research \\
  \texttt{corosset@microsoft.com} \\
   \And
  David Hawking \\
  \texttt{david.hawking@acm.org} \\
   \And
  Nick Craswell \\
  Microsoft AI \& Research \\
  \texttt{nickcr@microsoft.com} \\
   \And
  Fernando Diaz \\
  Microsoft AI \& Research \\
  \texttt{fdiaz@microsoft.com} \\
   \And
  Emine Yilmaz \\
  Microsoft AI \& Research \\
  \texttt{emine.yilmaz@ucl.ac.uk} \\
}

\begin{document}
\maketitle

\begin{abstract}
Classical information retrieval (IR) methods, such as query likelihood and BM25, score documents independently \wrt each query term, and then accumulate the scores.
Assuming \emph{query term independence} allows precomputing term-document scores using these models---which can be combined with specialized data structures, such as inverted index, for efficient retrieval.
Deep neural IR models, in contrast, compare the whole query to the document and are, therefore, typically employed only for late stage re-ranking.
We incorporate query term independence assumption into three state-of-the-art neural IR models: BERT, Duet, and CKNRM---and evaluate their performance on a passage ranking task.
Surprisingly, we observe no significant loss in result quality for Duet and CKNRM---and a small degradation in the case of BERT.
However, by operating on each query term independently, these otherwise computationally intensive models become amenable to offline precomputation---dramatically reducing the cost of query evaluations employing state-of-the-art neural ranking models.
This strategy makes it practical to use deep models for retrieval from large collections---and not restrict their usage to late stage re-ranking.
\end{abstract}


\section{Introduction}
\label{sec:intro}

Many traditional information retrieval (IR) ranking functions---\eg, \citep{robertson2009probabilistic, ponte1998language}---manifest the \emph{query-term independence} property---\ie, the documents can be scored independently \wrt each query term, and then the scores accumulated.
Given a document collection, these term-document scores can be precomputed and combined with specialized IR data structures, such as inverted indexes \citep{zobel2006inverted}, and clever organization strategies (\eg, impact-ordering \citep{anh2001vector}) to aggressively prune the set of documents that need to be assessed per query.
This dramatically speeds up query evaluations enabling fast retrieval from large collections, containing billions of documents.

Recent deep neural architectures---such as BERT \citep{nogueira2019passage}, Duet \citep{mitra2017learning}, and CKNRM \citep{dai2018convolutional}---have demonstrated state-of-the-art performance on several IR tasks.
However, the superior retrieval effectiveness comes at the cost of evaluating deep models with tens of millions to hundreds of millions of parameters at query evaluation time.
In practice, this limits the scope of these models to late stage re-ranking.
Like traditional IR models, we can incorporate the query term independence assumption into the design of the deep neural model---which would allow offline precomputation of all term-document scores.
The query evaluation then involves only their linear combination---alleviating the need to run the computation intensive deep model at query evaluation time.
We can further combine these precomputed machine-learned relevance estimates with an inverted index, to retrieve from the full collection.
This significantly increases the scope of potential impact of neural methods in the retrieval process.
We study this approach in this work.

Of course, by operating independently per query term, the ranking model has access to less information compared to if it has the context of the full query.
Therefore, we expect the ranking model to show some loss in retrieval effectiveness under this assumption.
However, we trade this off with the expected gains in efficiency of query evaluations and the ability to retrieve, and not just re-rank, using these state-of-the-art deep neural models.

In this preliminary study, we incorporate the query term independence assumption into three state-of-the-art neural ranking models---BERT \citep{nogueira2019passage}, Duet \citep{mitra2017learning}, and CKNRM \citep{dai2018convolutional}---and evaluate their effectiveness on the MS MARCO passage ranking task \citep{bajaj2016ms}.
We surprisingly find that the two of the models suffer no statistically significant adverse affect \wrt ranking effectiveness on this task under the query term independence assumption.
While the performance of BERT degrades under the strong query term independence assumption---the drop in MRR is reasonably small and the model maintains a significant performance gap compared to other non-BERT based approaches.
We conclude that at least for a certain class of existing neural IR models, incorporating query term independence assumption may result in significant efficiency gains in query evaluation at minimal (or no) cost to retrieval effectiveness.

\section{Related work}
\label{sec:related}

Several neural IR methods---\eg, \citep{ganguly2015word, kenter15short, nalisnick2016improving, guo2016deep}---already operate under query term independence assumption.
However, recent performance breakthroughs on many IR tasks have been achieved by neural models \citep{hu2014convolutional, pang2016text, mitra2017learning, dai2018convolutional, nogueira2019passage} that learn latent representations of the query or inspect interaction patterns between query and document terms.
In this work, we demonstrate the potential to incorporate query term independence assumption in these recent representation learning and interaction focused models.

Some neural IR models \citep{huang2013learning, gao2011clickthrough} learn low dimensional dense vector representations of query and document that can be computed independently during inference.
These models are also amenable to precomputation of document representations---and fast retrieval using approximate nearest neighbor search \citep{aumuller2017ann, boytsov2016off}.
An alternative involves learning higher dimensional but sparse representations of query and document \citep{salakhutdinov2009semantic, zamani2018neural2} that can also be employed for fast lookup.
However, these approaches---where the document representation is computed independently of the query---do not allow for interactions between the query term and document representations.
Early interaction between query and document representation is important to many neural architectures \citep{hu2014convolutional, pang2016text, mitra2017learning, dai2018convolutional, nogueira2019passage}.
The approach proposed in this study allows for interactions between individual query terms and documents.

Finally, we refer the reader to \citep{mitra2018introduction} for a more general survey of neural methods for IR tasks.

\section{Neural Ranking Models with Query Term Independence Assumption}
\label{sec:model}

IR functions that assume query term independence observe the following general form:

\begin{align}
    S_{q,d} &= \sum_{t \in q}{s_{t,d}}
\end{align}

Where, $s \in \mathbb{R}^{|V| \times |C|}_{\geq 0}$ is the set of positive real-valued scores as estimated by the relevance  model corresponding to documents $d \in C$ in collection $C$ \wrt to terms $t \in V$ in vocabulary $V$---and $S_{q,d}$ denotes the aggregated score of document $d$ \wrt to query $q$.
For example, in case of BM25 \citep{robertson2009probabilistic}:

\begin{align}
    s_{t,d} &= \text{idf}_t\cdot {\frac {\text{tf}_{td} \cdot (k_1+1)}{\text{tf}_{td}+k_1 \cdot \left(1-b+b \cdot {\frac{|d|}{avgdl}}\right)}}
\end{align}

Where, $\text{tf}$ and $\text{idf}$ denote term-frequency and inverse document frequency, respectively---and $k_1$ and $b$ are the free parameters of the BM25 model.

Deep neural models for ranking, in contrast, do not typically assume query term independence.
Instead, they learn complex matching functions to compare the candidate document to the full query.
The parameters of such a model $\phi$ is typically learned discriminatively by minimizing a loss function of the following form:

\begin{align}
    \mathcal{L} = \mathbb{E}_{q \sim \theta_q,\; d_{+} \sim \theta_{d_+}, d_{-} \sim \theta_{d_-}} [\ell(\Delta_{q, d_+, d_-})]
    \label{eqn:loss} \\
    \text{where,}\quad \Delta_{q, d_+, d_-} = \phi_{q, d_+} - \phi_{q, d_-}
\end{align}

We use $d_+$ and $d_-$ to denote a pair of relevant and non-relevant documents, respectively, \wrt query $q$.
The instance loss $\ell$ in Equation \ref{eqn:loss} can take different forms---\eg, ranknet \citep{burges2005learning} or hinge \citep{herbrich2000large} loss.

\begin{align}
    \ell_\text{ranknet}(\Delta_{q, d_+, d_-}) &= \text{log}(1 + e^{-\sigma\cdot\Delta_{q, d_+, d_-}}) \\
    \ell_\text{hinge}(\Delta_{q, d_+, d_-}) &= \text{max}\{0, \epsilon - \Delta_{q, d_+, d_-}\}
\end{align}

Given a neural ranking model $\phi$, we define $\Phi$---the corresponding model under the query term independence assumption---as:

\begin{align}
    \Phi_{q,d} &= \sum_{t \in q}{\phi_{t,d}}
\end{align}

The new model $\Phi$ preserves the same architecture as $\phi$ but estimates the relevance of a document independently \wrt each query term.
The parameters of $\Phi$ are learned using the modified loss:

\begin{align}
    \mathcal{L} = \mathbb{E}_{q \sim \theta_q,\; d_{+} \sim \theta_{d_+}, d_{-} \sim \theta_{d_-}} [\ell(\delta_{q, d_+, d_-})]
    \label{eqn:loss-new} \\
    \text{where,}\quad \delta_{q, d_+, d_-} = \sum_{t \in q}{\phi_{t, d_+} - \phi_{t, d_-}}
\end{align}

Given collection $C$ and vocabulary $V$, we precompute $\phi_{t,d}$ for all $ t \in V$ and $d \in C$.
In practice, the total number of combinations of $t$ and $d$ may be large but we can enforce additional constraints on which $\langle t, d \rangle$ pairs to evaluate, and assume no contributions from remaining pairs.
During query evaluation, we can lookup the precomputed score $\phi_{t,d}$ without dedicating any additional time and resource to evaluate the deep ranking model.
We employ an inverted index, in combination with the precomputed scores, to perform retrieval from the full collection using the learned relevance function $\Phi$.
We note that several IR data structures assume that $\phi_{t,d}$ be always positive which may not hold for any arbitrary neural architecture.
But this can be addressed by applying a rectified linear unit activation on the model's output.
The remainder of this paper describes our empirical study and summarizes our findings.

\section{Experiments}
\label{sec:method}

\subsection{Task description}
\label{sec:method-task}

We study the effect of the query term independence assumption on deep neural IR models in the context of the MS MARCO passage ranking task \citep{bajaj2016ms}.
We find this ranking task to be suitable for this study for several reasons.
Firstly, with one million question queries sampled from Bing's search logs, 8.8 million passages extracted from web documents, and 400,000 positively labeled query-passage pairs for training, it is one of the few large datasets available today for benchmarking deep neural IR methods.
Secondly, the challenge leaderboard\footnote{\url{http://www.msmarco.org/leaders.aspx}}---with 18 entries as of March 3, 2019---is a useful catalog of approaches that show state-of-the-art performance on this task.
Conveniently, several of these high-performing models include public implementations for the ease of reproducibility.


The MS MARCO passage ranking task comprises of one thousand passages per query that the IR model, being evaluated, should re-rank.
Corresponding to every query, one or few passages have been annotated by human editors as containing the answer relevant to the query.
The rank list produced by the model is evaluated using the mean reciprocal rank (MRR) metric against the ground truth annotations.
We use the MS MARCO training dataset to train all baseline and treatment models, and report their performance on the publicly available development set which we consider---and hereafter refer to---as the test set for our experiments.
This test set contains about seven thousand queries which we posit is sufficient for reliable hypothesis testing.

Note that the thousand passages per query were originally retrieved using BM25 from a collection that is provided as part of the MS MARCO dataset.
This allows us to also use this dataset in a retrieval setting---in addition to the re-ranking setting used for the official challenge.
We take advantage of this in our study.

\subsection{Baseline models}
\label{sec:method-baseline}

We begin by identifying models listed on the MS MARCO leaderboard that can serve as baselines for our work.
We only consider the models with public implementations.
We find that a number of top performing entries---\eg, \citep{nogueira2019passage}---are based on recently released large scale language model called BERT \citep{devlin2018bert}.
The BERT based entries are followed in ranking by the Duet \citep{mitra2017learning} and the Convolutional Kernel-based Neural Ranking Model (CKNRM) \citep{dai2018convolutional}.
Therefore, we limit this study to BERT, Duet, and CKNRM.

\paragraph{BERT}
\label{sec:method-baseline-bert}
\citet{nogueira2019passage} report state-of-the-art retrieval performance on the MS MARCO passage re-ranking task by fine tuning BERT \citep{devlin2018bert} pretrained models.
In this study, we reproduce the results from their paper corresponding to the BERT Base model and use it as our baseline.
Under the term independence assumption, we evaluate the BERT model once per query term---wherein we input the query term as sentence A and the passage as sentence B.

\begin{table}[t]
    \centering
    \caption{Comparing ranking effectiveness of BERT, Duet, and CKNRM with the query independence assumption (denoted as ``Term ind.'') with their original counterparts (denoted as ``Full'').
    The difference between the median MRR for ``full'' and ``term ind.'' models are \emph{not} statistically significant based on a student's t-test $(p < 0.05)$ for Duet and CKNRM.
    The difference in MRR is statistically significant based on a student's t-test $(p < 0.05)$ for BERT (single run).
    The BM25 baseline (single run) is included for reference.}
    \begin{tabular}{lrlll}
    \hline
    \hline
        \multirow{2}{*}{\textbf{Model}} & \multicolumn{4}{c}{\textbf{MRR@10}} \\
        & Mean & ($\pm$ Std. dev) & & Median \\
        \hline
        \multicolumn{5}{l}{\textbf{BERT}} \\
        Full & $0.356$ & & & $0.356$ \\
        Term ind. & $\mathbf{0.333}$ & & & $\mathbf{0.333}$ \\
        \hline
        \multicolumn{5}{l}{\textbf{Duet}} \\
        Full & $0.239$ & $(\pm 0.002)$ & & $0.240$ \\
        Term ind. & $0.244$ & $(\pm 0.002)$ & & $0.244$ \\
        \hline
        \multicolumn{5}{l}{\textbf{CKNRM}} \\
        Full & $0.223$ & $(\pm 0.004)$ & & $0.224$ \\
        Term ind. & $0.222$ & $(\pm 0.005)$ & & $0.221$ \\
        \hline
        BM25 & $0.167$ & & & $0.167$ \\
        \hline
        \hline
    \end{tabular}
    \label{tbl:results-mrr}
\end{table}

\paragraph{Duet}
\label{sec:method-baseline-duet}
The Duet \citep{mitra2017learning} model estimates the relevance of a passage to a query by a combination of
\begin{enumerate*}[label=(\roman*)]
    \item examining the patterns of exact matches of query terms in the passage, and
    \item computing similarity between learned latent representations of query and passage.
\end{enumerate*}
Duet has previously demonstrated state-of-the-art performance on TREC CAR \citep{nanni2017benchmark} and is an official baseline for the MS MARCO challenge.
The particular implementation of Duet listed on the leaderboard includes modifications\footnote{\url{https://github.com/dfcf93/MSMARCO/blob/master/Ranking/Baselines/Duet.ipynb}} to the original model \citep{mitra2019updated}.
We use this provided implementation for our study.
Besides evaluating the model once per query term, no additional changes were necessary to its architecture under the query term independence assumption.

\paragraph{CKNRM}
\label{sec:method-baseline-cknrm}
The CKNRM model combines kernel pooling based soft matching \citep{xiong2017end} with a convolutional architecture for comparing $n$-grams.
CKNRM uses kernel pooling to extract ranking signals from interaction matrices of query and passage $n$-grams.
Under the query term independence assumption, the model considers one query term at a time---and therefore we only consider the interactions between the query unigrams and passage $n$-grams.
We base our study on the public implementation\footnote{\url{https://github.com/thunlp/Kernel-Based-Neural-Ranking-Models}} of this model.

\bigskip\noindent
For all models we re-use the published hyperparameter values and other settings from the MS MARCO website.

\section{Results}
\label{sec:result}

\begin{table}[t]
    \centering
    \caption{Comparing Duet (with query term independence assumption) and BM25 under the full retrieval settings on a subset of MS MARCO dev queries.
    The differences in recall and MRR between Duet (term ind.) and BM25 are statistically significant according to student's t-test $(p < 0.01)$.}
    \begin{tabular}{lcc}
    \hline
    \hline
        \textbf{Model} & \textbf{Recall@1000}  & \textbf{MRR@10} \\
        \hline
        BM25 & $0.80$ & $0.169$ \\
        Duet (term ind.) & $\mathbf{0.85}$ & $\mathbf{0.218}$ \\
        \hline
        \hline
    \end{tabular}
    \label{tbl:results-recall}
\end{table}

Table \ref{tbl:results-mrr} compares the BERT, the Duet, and the CKNRM models trained under the query term independence assumption to their original counterparts on the passage re-ranking task.
We train and evaluate the Duet and the CKNRM based models five and eight times, respectively, using different random seeds---and report mean and median MRR.
For the BERT based models, due to long training time we only report results based on a single training and evaluation run.
As table \ref{tbl:results-mrr} shows, we observe no statistically significant difference in effectiveness from incorporating the query term independence assumptions in either Duet or CKNRM.
The query term independent BERT model performs slightly worse than its original counterpart on MRR but the performance is still superior to other non-BERT based approaches listed on the public leaderboard.

We posit that models with query term independence assumption---even when slightly less effective compared to their full counterparts---are likely to retrieve better candidate sets for re-ranking.
To substantiate this claim, we conduct a small-scale retrieval experiment based on a random sample of 395 queries from the test set.
We use the Duet model with the query term independence assumption to precompute the term-passage scores constrained to 
\begin{enumerate*}[label=(\roman*)]
    \item the term appears at least once in the passage, and
    \item the term does not appear in more than $5\%$ of the passage collection.
\end{enumerate*}
Table \ref{tbl:results-recall} compares Duet and BM25 on their effectiveness as a first stage retrieval method in a potential telescoping setting \citep{matveeva2006high}.
We observe a $6.25\%$ improvement in recall@1000 from Duet over the BM25 baseline.
To perform similar retrieval from the full collection using the full Duet model, unlike its query-term-independent counterpart, is prohibitive because it involves evaluating the model on every passage in the collection against every incoming query.

\section{Discussion and conclusion}
\label{sec:conclusion}

The emergence of compute intensive ranking models, such as BERT, motivates rethinking how these models should be evaluated in large scale IR systems.
The approach proposed in this paper moves the burden of model evaluation from the query evaluation stage to the document indexing stage.
This may have further consequences on computational efficiency by allowing batched model evaluation that more effectively leverages GPU (or TPU) parallelization.

This preliminary study is based on three state-of-the-art deep neural models on a public passage ranking benchmark.
The original design of all three models---BERT, Duet, and CKNRM---emphasize on early interactions between query and passage representations.
However, we observe that limiting the interactions to passage and individual query terms has reasonably small impact on their effectiveness.
These results are promising as they support the possibility of dramatically speeding up query evaluation for some deep neural models, and even employing them to retrieve from the full collection.
The ability to retrieve---and not just re-rank---using deep models has significant implications for neural IR research.
Any loss in retrieval effectiveness due to incorporating strong query term independence assumptions may be further recovered by additional stages of re-ranking in a telescoping approach \citep{matveeva2006high}.

This study is focused on the passage ranking task.
The trade-off between effectiveness and efficiency may be different for document retrieval and other IR tasks.
Traditional IR methods in more complex retrieval settings---\eg, when the document is represented by multiple fields \citep{robertson2004simple}---also observe the query term independence assumption.
So, studying the query term independence assumption in the context of corresponding neural models---\eg, \citep{zamani2018neural}---may also be appropriate.
We note these as important future directions for our research.

The findings from this study may also be interpreted as pointing to a gap in our current state-of-the-art neural IR models that do not take adequate advantage of term proximity signals for matching.
This is another finding that may hold interesting clues for IR researchers who want to extract more retrieval effectiveness from deep neural methods.

\bibliographystyle{plainnat}
\bibliography{bibtex}

\end{document}